\documentclass[twocolumn,showpacs,preprintnumbers,amsmath,amssymb,floatfix]{revtex4}
\usepackage{graphicx}
\usepackage{dcolumn}
\usepackage{bm}
\usepackage[usenames]{color}

\begin{document}

\title{Linear Algebraic Calculation of Green's function \\
for Large-Scale Electronic Structure Theory}
\author{R. Takayama$^{1,2}$}
    \altaffiliation[Present address:]{Canon Inc., Analysis Technology Center, 
30-2 Shimomaruko 3-chome, Ohta-ku, Tokyo 146-8501, Japan}    
\author{T. Hoshi$^{2,3}$}
\author{T. Sogabe$^{2}$}
    \altaffiliation[Present address:]{Department of Computational Science and Engineering, Nagoya University, Furo-cho, Chikusa-ku, Nagoya 464-8603, Japan}    
\author{S.-L. Zhang$^{2}$}
    \altaffiliation[Present address:]{Department of Computational Science and Engineering, Nagoya University, Furo-cho, Chikusa-ku, Nagoya 464-8603, Japan}    
\author{T. Fujiwara$^{2,3}$}
    \affiliation{$^{1}$ Research and Development for Applying Advanced Computational Science and Technology (ACT-JST), Japan Science and Technology Agency, 4-1-8 Honcho, Kawaguchi-shi, Saitama 332-0012, Japan} 
    \affiliation{$^{2}$ Department of Applied Physics, University of Tokyo,7-3-1 Hongo, Bunkyo-ku, Tokyo 113-8656, Japan}
    \affiliation{$^{3}$ Core Research for Evolutional Science and Technology  (CREST-JST), Japan Science and Technology Agency, 4-1-8 Honcho, Kawaguchi-shi, Saitama 332-0012, Japan}
\date{\today}
\begin{abstract}
A linear algebraic method named the shifted conjugate-orthogonal-conjugate-gradient  
method is introduced for large-scale electronic structure calculation.
The method gives
an iterative solver algorithm of 
the Green's function and the density matrix 
without calculating eigenstates.
The problem is reduced to 
independent linear equations at many energy points
and the calculation is actually carried out 
only for a single energy point. 
The method is robust against the round-off error and 
the calculation can reach the machine accuracy.
With the observation of residual vectors,
the accuracy can be controlled, 
microscopically, independently for each element of the Green's function, and
dynamically, at each step in dynamical simulations.
The method is applied to both semiconductor and metal.
\end{abstract}
\maketitle

\section{Introduction}
\label{SECTION-INTRO}

Large-scale atomistic simulation with quantum mechanical freedom of electrons 
requires manipulation of a large Hamiltonian matrix. 
In order to calculate physical quantities of a system, 
we should obtain either eigenstates or the density matrix of the system. 
The calculation of eigenstates is usually reduced to 
matrix diagonalization procedure and 
this procedure results in severe computational cost for a large-scale system. 

Any physical quantity $X$ can be evaluated 
by means of the density matrix $\rho$ as 
\begin{equation}
\langle X \rangle 
  = \int\int d{\bm r}d{\bm r}^\prime \rho ({\bm r},{\bm r}^\prime ) X({\bm r}^\prime, {\bm r}) .
\label{eq:physica-quantity}
\end{equation}
Even though the density matrix is of long-range, only the short-range behavior 
of the density matrix is necessary, 
if $X$ is a  short-range  operator.
The energy and forces acting on an individual atom are really this case,  
and the locality of the Hamiltonian realizes this feature 
in large-scale calculation. 
Moreover, the density matrix $\rho$ can be obtained 
from the Green's function.
Therefore, the essential methodology for large-scale electronic 
structure calculation and molecular dynamics (MD) simulation 
is how to obtain density matrix $\rho$ 
or Green's function without calculating eigenstates.
\cite{Kohn96,Haydock,Galli,Wu-Jayanthi,Roche-Mayou, Roche2005, 
Ozaki-Terakura2001, OKAndersen.00,
hoshi-fujiwara.00,hoshi-fujiwara.01,hoshi-fujiwara.03,
hoshi.03,geshi_etal.03,takayama_etal.04,hoshi-fujiwara.05}

We have developed a set of methods for large-scale atomistic simulation 
without calculating eigenstates in fully quantum mechanical 
description of electron systems.~\cite{hoshi-fujiwara.00,hoshi-fujiwara.01,hoshi-fujiwara.03,hoshi.03,geshi_etal.03,takayama_etal.04,hoshi-fujiwara.05} 
Among them, the subspace diagonalization method based 
on Krylov subspace (SD-KS method) was introduced,
where the original Hamiltonian matrix $H$ is reduced to a small size easy-tractable one and  
its diagonalization leads to  approximation of the density matrix of 
the original system.~\cite{takayama_etal.04} 
The first important feature of the SD-KS method is that 
we can monitor numerical accuracy  during the simulation
using a residual error of the Green's function,  
as shown in the present paper. 
The second important feature is  that the SD-KS method can be used 
both for  metallic and insulating systems. 
We found, however, 
that the SD-KS method has a numerical instability, when the energy spectrum is 
calculated with a {\it very fine} energy resolution, 
as discussed in Appendix \ref{sec:appendix}. 

Then our new strategy to obtain the  Green's function and the density matrix 
is to solve linear equations 
with  a given  basis $|j \rangle$; 
\begin{equation}
(z-H)|x_j \rangle = |j \rangle .
\label{eq:linear-eq0}
\end{equation}
In the present case, the Hamiltonian $H$ is real symmetric 
and $(z-H)$ is not Hermitian  but complex symmetric 
with a complex energy ($z \equiv E + i \gamma$).  
Once the linear equation is solved, 
one can obtain any element of the Green's function as 
\begin{equation}
G_{ij}(z) =  \langle i |(z-{H})^{-1} |j \rangle = \langle i |x_j \rangle. 
\label{eq:Green}
\end{equation}
The numerical energy integration 
is required to obtain the one-body density matrix;
\begin{equation}
 \rho_{ij} = -\frac{1}{\pi} \int_{-\infty}^{\infty}
 {\rm Im} \, G_{ij}(E + i \gamma) \,
 f \left(\frac{E - \mu}{k_{\rm B}\tau} \right) {\rm d}E,
 \label{eq:rho_int}
\end{equation}
with  the Fermi distribution function $f(x)$ 
and a small imaginary part of energy $(\gamma \rightarrow  0+)$.
The chemical potential $\mu$ is  determined so that 
the sum of the diagonal elements of the density matrix equals  
the total number of electrons.

The aim of the present paper is to introduce 
the shifted conjugate-orthogonal conjugate-gradient (COCG) method,
an iterative solver algorithm of eq.~(\ref{eq:linear-eq0}). 
The Green's function and the density matrix are obtained 
using eqs.~(\ref{eq:Green}) and (\ref{eq:rho_int}).  
The shifted COCG method shares the before-mentioned two features 
with the SD-KS method.
Moreover, 
the third important feature of the shifted COCG method, different from the 
SD-KS method, is the robustness against round-off error 
and the calculation can reach the machine accuracy. 

The present paper is organized as follows;
In Sec.~\ref{sec:sCOCG},  Krylov subspace  will be explained 
and the shifted COCG method will be introduced. 
In Sec.~\ref{RN_and_Opera}, 
the residual norm (RN) will be introduced  
to monitor the convergence behavior of the method.
The number of operations in actual calculation is also discussed. 
Section~\ref{sec:example} is devoted to application 
of the present method to 
an atomic scale reconstruction of semiconductor surface (silicon) and 
the bulk electronic structure of metal (copper).
The conclusion will be given
in Sec.~\ref{sec:conclusion}. 
In Appendix \ref{sec:appendix}, 
several numerical aspects will be discussed 
for the shifted COCG and the SD-KS methods and 
the difference between the two method will be clarified.

\section{Shifted conjugate-orthogonal conjugate-gradient method}
\label{sec:sCOCG}
\subsection{Shifted systems and Krylov subspace}
\label{sec:KrylovSS}

Now we should concentrate the method to solve eq.~(\ref{eq:linear-eq0}) 
with a large matrix $H$ and a fixed basis $| j \rangle$. 
The method should be iterative and not require 
the matrix inversion procedure of $(z-H)$. 
The problem is reduced to the linear equations,
eq.~(\ref{eq:linear-eq0}), for a given set 
of energy points $z = z_1, z_2,z_3....$. 
These linear equations are called 
\lq shifted' linear equations or 
\lq shifted' linear systems 
in mathematical textbooks, 
because {\it shifted} matrices $(z_1-H)$,  $(z_2-H)$,  $(z_3-H)$,.. appear.
If the equations are solved independently among different energy points,
the total computational cost is proportional 
to the number of energy points $N_{\rm ene}$.
The essence of the present method is 
that we should solve the equation 
{\it only at one energy point} and the  solutions 
at other energy points are given 
with a moderate computational cost.

The present method is realized 
using Krylov subspace (KS).~\cite{saad.00,Vorst.03}
Krylov subspace is defined 
for an arbitrary matrix $A$ and vector $|j \rangle$,   
as the linear space spanned by  
a set of states (vectors) $\{A^n |j\rangle\}$;
\begin{equation}
\label{eq:krylov1}
K_{n_{}}(A, |j \rangle) \equiv {\rm span}\left\{
|j \rangle, \  A|j \rangle, \  A^2|j \rangle, \  \ldots, \  A^{n_{}-1}|j  \rangle
\right\} ,
\end{equation}
where $n$ is the dimension of the KS. 
Iterative methods based on KS,
such as the standard conjugate-gradient algorithm,
are generally called Krylov subspace methods. 
In the present method, 
the solution vector $|x_j\rangle$ of eq.~(\ref{eq:linear-eq0}), 
is constructed within the KS of  
$K_n (z-H, |j  \rangle )$
at the $n$-th iteration. 

The present method, the shifted COCG method,
is a combined method of two KS algorithms;
(a) the conjugate-orthogonal conjugate-gradient method (COCG method) 
~\cite{Vorst-Melissen.93} and 
(b) the theorem of collinear residual for shifted linear systems
~\cite{Frommer.02}. 
The essential point is  that 
the KS among shifted systems gives the same linear space 
\begin{equation}
K_n (z_1-H, |j  \rangle ) = K_n (z_2-H, |j  \rangle ).
\label{EQ-SHIFTED-KRYLOV}
\end{equation}
The actual procedures are given in the next subsection.

\subsection{Shifted COCG method} 
\label{sec:shifted_COCG}

Here we present the formulation of the shifted COCG algorithm,
following Ref.~\cite{Frommer.02}; 
We pick out arbitrarily one energy point as 
\lq reference' energy point 
$z_{\rm ref}\equiv E_{\rm ref} + i \gamma$.
Equation ~(\ref{eq:linear-eq0}) 
at the reference energy ($z=z_{\rm ref}$) 
is reformulated as
\begin{eqnarray}
 A \bm{x} = \bm{b}, 
  \label{EQ-REF}
\end{eqnarray}
where the matrix $A$ is defined as $A \equiv z_{\rm ref} -H$ and 
the suffix $j$ is dropped 
($| j \rangle \Rightarrow \bm{b},  | x_j \rangle \Rightarrow \bm{x}$). 
Since the matrix $A$ is not Hermitian, 
the matrix-vector notation is used in this subsection,
rather than the bracket notation.
Hereafter the equation at $z=z_{\rm ref}$  
is called \lq reference'  system.

For eq.~(\ref{EQ-REF}),
we use the COCG algorithm ~\cite{Vorst-Melissen.93}, 
a standard iterative algorithm for a linear equation
with a complex symmetric matrix $A$. ~\cite{Note_Frommer}
At the $n$-th iteration, 
the solution vector $\bm{x}_n$, 
the residual vector $\bm{r}_n$, and 
the search direction vector $\bm{p}_n$ are represented as 
\begin{eqnarray}
\bm{x}_n &=& \bm{x}_{n-1} + \alpha_{n-1} \bm{p}_{n-1},   \label{eq:xn_def} \\
\bm{r}_n &=& \bm{r}_{n-1} - \alpha_{n-1} A \bm{p}_{n-1}, \label{eq:rn_def} \\
\bm{p}_n &=& \bm{r}_n + \beta_{n-1} \bm{p}_{n-1},        \label{eq:pn_def}
\end{eqnarray}
respectively. 
Here, coefficients $\alpha_{n}$ and $\beta_{n}$ are given as 
\begin{eqnarray}
\alpha_{n} &=& \frac{\bm{r}_n^{\rm T} \bm{r}_n}{\bm{p}_n^{\rm T} A\bm{p}_n}, 
\label{eq:def_alpha} \\ 
\beta_{n}  &=& \frac{\bm{r}_{n+1}^{\rm T} \bm{r}_{n+1}}{\bm{r}_{n}^{\rm T} \bm{r}_{n}}. 
\label{eq:def_beta} 
\end{eqnarray}
The initial conditions for iteration are 
${\bm x}_0={\bm p}_{-1}=0$, 
${\bm r}_0=\bm{b}$, 
$\beta_{-1}=0$,  
$\alpha_{-1}=1$.
Note here that the inner products are given as 
$\displaystyle{\bm{a}^{\rm T} \bm{b} \, (\neq \bm{a}^{\rm H} \bm{b} })$.
When the iteration number $n$ reaches the matrix dimension of $A$,
denoted $M$,
the residual vector should be zero
and the solution vector should be exact 
($\bm{r}_M = 0, \bm{x}_M = A^{-1}\bm{b}$).

Eliminating $\bm{p}_{n-1}$ from eq.~(\ref{eq:rn_def}) with eq.~(\ref{eq:pn_def}), 
we obtain three-term recurrence relation for the residual vector;
\begin{equation}
\bm{r}_{n+1} = -\alpha_n A \bm{r}_n 
+ \left( 1+ \frac{\beta_{n-1} \alpha_{n}}{\alpha_{n-1}} \right) \bm{r}_n
-\frac{\beta_{n-1} \alpha_{n}}{\alpha_{n-1}} \bm{r}_{n-1} 
\label{eq:residual_seed} .
\end{equation}
The most time-consuming part of the COCG algorithm 
is the matrix-vector product ($A\bm{p}_n$) in 
eq.~(\ref{eq:def_alpha}). 
This matrix-vector product corresponds to 
the procedure for updating the KS; 
$K_n(A, \bm{b}) \Rightarrow K_{n+1}(A, \bm{b})$.

Similarly, 
we reformulate eq.(\ref{eq:linear-eq0}) 
with a shifted energy point $z = z_{\rm ref} + \sigma $ as
\begin{eqnarray}
 (A + \sigma I ) \bm{x} = \bm{b}. 
  \label{EQ-SHIFT}
\end{eqnarray}
For the \lq shifted' system, 
the $n$-th solution vector $\bm{x}_n^{\sigma}$ and 
the search direction vector $\bm{p}_n^{\sigma}$ are given as 
\begin{eqnarray}
\bm{x}_n^{\sigma} &=& \bm{x}_{n-1}^{\sigma} + \alpha_{n-1}^{\sigma} \bm{p}_{n-1}^{\sigma}, 
\label{eq:xn_def_shift} \\
\bm{p}_n^{\sigma} &=& \bm{r}_n^{\sigma} + \beta_{n-1}^{\sigma} \bm{p}_{n-1}^{\sigma}.      
\label{eq:pn_def_shift}
\end{eqnarray}
The initial values of the vectors or coefficients are chosen 
to be the same as in the reference system.
The equation corresponding to eq.(\ref{eq:residual_seed}) of the shifted system is 
\begin{eqnarray}
 \bm{r}_{n+1}^{\sigma} &=& -\alpha_n^{\sigma} (A + \sigma I ) \bm{r}_n^{\sigma} \nonumber \\
& & + \left( 1+ \frac{\beta_{n-1}^{\sigma} \alpha_{n}^{\sigma}}{\alpha_{n-1}^{\sigma}} \right) 
\bm{r}_n^{\sigma} 
-\frac{\beta_{n-1}^{\sigma} \alpha_{n}^{\sigma}}{\alpha_{n-1}^{\sigma}} \bm{r}_{n-1}^{\sigma} .
\label{eq:residual_shift}
\end{eqnarray}

Since the KS between the reference and 
shifted systems are equivalent
($K_n(A, \bm{b}) = K_n(A+\sigma I, \bm{b})$),
as stated in eq.~(\ref{EQ-SHIFTED-KRYLOV}),
one can prove 
that the residual vectors between them, 
$\bm{r}_n^{\sigma}$ and $\bm{r}_n$, 
are collinear;
\begin{equation}
\bm{r}_n^{\sigma} = \frac{1}{\pi_n^{\sigma}} \bm{r}_n. 
\label{eq:shift-res}
\end{equation}
which is 
the theorem of collinear residual for shifted linear systems.
\cite{Frommer.02}
With eq.~(\ref{eq:shift-res}), eq.(\ref{eq:residual_seed}) can be
modified as 
\begin{eqnarray}
\bm{r}_{n+1}^{\sigma} 
&=& -\frac{\pi_{n}^{\sigma}}{\pi_{n+1}^{\sigma}} \alpha_n (A + \sigma I)  \bm{r}_n^{\sigma} 
\nonumber \\
&&+ \frac{\pi_{n}^{\sigma}}{\pi_{n+1}^{\sigma}}
\left( 1 + \alpha_n \sigma+ \frac{\beta_{n-1} \alpha_{n}}{\alpha_{n-1}} \right) \bm{r}_n^{\sigma}
\nonumber \\
&&-\frac{\pi_{n-1}^{\sigma}}{\pi_{n+1}^{\sigma}}\frac{\beta_{n-1} \alpha_{n}}{\alpha_{n-1}} \bm{r}_{n-1}^{\sigma}.
\label{eq:residual_shift_0}
\end{eqnarray}
Comparing the coefficients in eqs.~(\ref{eq:residual_shift_0}) and
(\ref{eq:residual_shift}),
we obtain
\begin{eqnarray}
\alpha_n^{\sigma} &=& \frac{\pi_{n}^{\sigma}}{\pi_{n+1}^{\sigma}} \alpha_n , 
\label{eq:alpha_shift} \\
\beta_n^{\sigma} &=& \left( \frac{\pi_{n}^{\sigma}}{\pi_{n+1}^{\sigma}} \right)^2 \beta_n ,
\label{eq:beta_shift} \\
\pi_{n+1}^{\sigma} &=&
\left( 1+  \alpha_n \sigma +  \frac{\beta_{n-1} \alpha_{n}}{\alpha_{n-1}} \right) 
\pi_n^{\sigma} 
-\frac{\beta_{n-1} \alpha_{n}}{\alpha_{n-1}} \pi_{n-1}^{\sigma} . 
\nonumber \\ 
 & & 
\label{eq:coeff_shift00} 
\end{eqnarray}
with the initial values of  
$\pi_0^\sigma = \pi_{-1}^\sigma =1$.
We can update the vector 
$\bm{r}_n^{\sigma}$
and the coefficients 
$\alpha_n^{\sigma}$ and $\beta_n^{\sigma}$
using eqs.~(\ref{eq:shift-res}),
(\ref{eq:alpha_shift}),(\ref{eq:beta_shift}), and (\ref{eq:coeff_shift00}),
which {\it do not} include any matrix-vector product. 
Consequently, the time-consuming procedure of the matrix-vector product
is needed only for the reference system,
which reduces the computational cost drastically. 
The detail of the computational cost will be estimated
in Sec.~\ref{Number_of_operations}.

We note here that the shift parameter $\sigma (\equiv z - z_{\rm ref})$ 
is an arbitrary complex variable in principle but 
we choose the value to be real ($\sigma = E- E_{\rm ref}$)
in  all the practical calculations of the present paper.
We also note that 
a practical electronic-structure calculation 
can be parallelized  with the present method, 
because the original problem of
eq.~(\ref{eq:linear-eq0}) is independent 
with respect to the basis suffix $j$.

\section{Convergence behavior and computational cost}
\label{RN_and_Opera} 
\subsection{Convergence behavior with residual norm}
\label{SECTION-RN}

Since the shifted COCG method is an iterative solver algorithm 
for eq.~(\ref{eq:linear-eq0}), 
we should establish a systematic procedure 
to find an optimal iteration number
in the context of electronic structure calculation.
In this section, such a systematic procedure 
is introduced by monitoring the norm of residual vector.

At the $n$-th iteration,  
we denote the solution vectors for the reference and shifted systems as
$| x_{n}^{(j)} \rangle$ and 
$ | x_{n}^{\sigma (j)} \rangle$,
respectively. 
The corresponding residual vectors are written by
\begin{eqnarray}
| r_{n}^{(j)} \rangle &\equiv & | j \rangle - 
 (z_{\rm ref}  - H ) | x_{n}^{(j)} \rangle
\label{eq:residual_G0} \\
| r_{n}^{\sigma (j)} \rangle &\equiv & | j \rangle - 
 (z_{\rm ref} + \sigma - H ) | x_{n}^{\sigma (j)} \rangle
=\frac{1}{\pi_{n}^{\sigma}} | r_{n}^{(j)} \rangle, 
\label{eq:residual_G} 
\end{eqnarray}
respectively.
The last equality is given by eq.~(\ref{eq:shift-res}).

Since we need only the elements of 
the density matrix 
among near-sited orbital pairs
or of the short-distance components in eq.~(\ref{eq:physica-quantity}), 
the convergence is necessary only  
for these components,
but not for far-distance components.
Therefore, 
we define a residual norm (RN)
from the components
only among these near-sited orbitals ($| i \rangle$) 
that are determined by the interaction range of Hamiltonian;
\begin{eqnarray}
|| {\bm r}_n^{(j)} ||^2 
&\equiv& \sum_i^{M_{\rm int}} |\langle i| r_n^{(j)} \rangle |^2 ,
\label{eq:ref-p-RN} \\
|| {\bm r}_n^{\sigma (j)} ||^2 
&\equiv& \sum_i^{M_{\rm int}} |\langle i| r_n^{\sigma (j)} \rangle |^2 
= \left| \frac{1}{\pi_{n}^{\sigma}} \right|^2 \, || {\bm r}_n^{(j)} ||^2,
\label{eq:shif-p-RN}
\end{eqnarray}
where $M_{\rm int}$ is  the number of interacting orbitals $|i\rangle$ 
for a basis $| j \rangle$, typically $10$ to $10^2$.
Note that the RN is an energy-dependent quantity. 

Since the Green's function should be integrated 
over  the given set of energy points to obtain the density matrix, 
we need to know the convergence behavior of the RN 
over the entire energy range.
Then we average the RN over 
the energy range ($E_{\rm min} < E<    E_{\rm max}$
or $E_{\rm min} -E_{\rm ref}<  \sigma <    E_{\rm max}-E_{\rm ref}$).
We call the resultant quantity the energy-averaged residual norm (a-RN);
\begin{eqnarray}
R_{n}^{(j)} & \equiv & 
\frac{1}{E_{\rm max} - E_{\rm min}} \int_{E_{\rm min}}^{E_{\rm max}} dE
|| {\bm r}_n^{\sigma (j)} ||^2   \nonumber \\
&=& \xi_{n}^{(j)} || {\bm r}_n^{(j)} ||^2,  
\label{eq:conv_fact00} 
\end{eqnarray}
where
\begin{eqnarray}
\xi_{n}^{(j)} &=& \frac{1}{E_{\rm max} - E_{\rm min}}
  \int_{E_{\rm min}}^{E_{\rm max}} 
  \left| \frac{1}{\pi_{n}^{\sigma}} \right|^2 dE.
\label{eq:cf-sum}
\end{eqnarray}
Since the a-RN, $R_{n}^{(j)}$, 
 can be monitored at every iteration ($n$), 
the optimal number of iterations can be determined 
by the value of $R_{n}^{(j)}$. 
The above determination
is carried out 
for the {\it microscopic} freedoms or  
individually among the basis suffix $j$.

\begin{figure}[htbp] 
\begin{center}
\resizebox{0.65\textwidth}{!}{
  \includegraphics[height=18cm,clip]{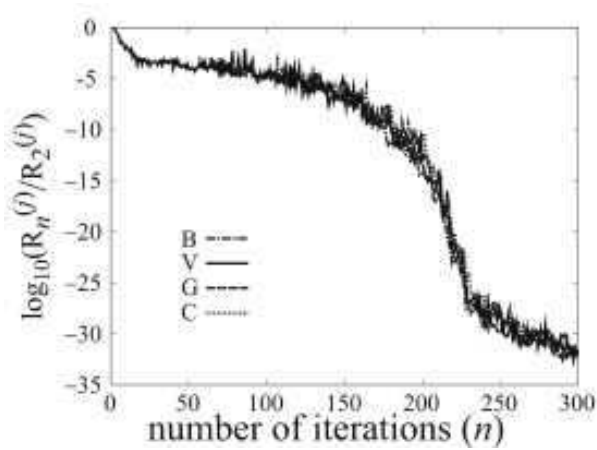}
}
\end{center}
\vspace{-5mm}
\caption{\label{fig:sCOCG_totalCF} 
Decay behavior of the a-RN for Si crystal with 512 atoms. 
A universal curve appears 
with four different reference energy points
that are placed
at the band bottom (B), 
in the valence band (V), 
in the gap (G) and in the conduction band (C).
See Fig.~\ref{fig:cocgs01ldos0}(a) 
for the actual positions of these energy points.
}
\end{figure}
\begin{figure}[htbp] 
\begin{center}
 \includegraphics[height=6cm,clip]{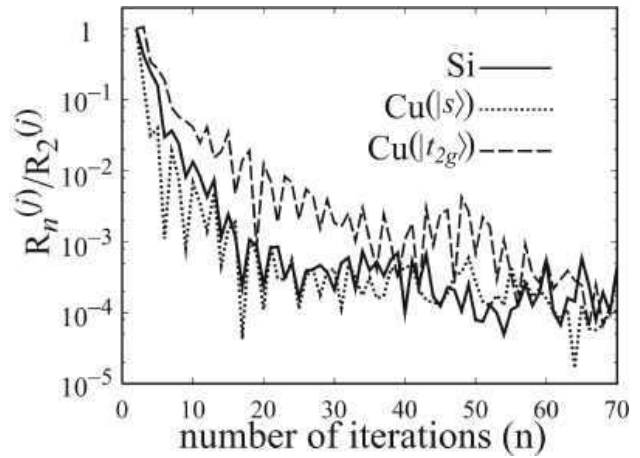}
\end{center}
\caption{\label{fig:total_vs_local}
Early stage in the decay behavior of the a-RN  
for Si crystal with 512 atoms 
and fcc Cu with 1568 atoms.
In the Cu system,
the two cases are plotted
with the starting vectors $|j \rangle$ 
of s and  t$_{2g}$ bases on an atom.  
}
\end{figure}

%
As an example, we calculated the electronic structure in  
a bulk Si system of 512 atoms with a cubic simulation cell. 
We use a transferable Hamiltonian of silicon 
in the Slater-Koster form 
of s and p orbitals ~\cite{kwon-etal.94}.  
The matrix dimension of the Hamiltonian $H$ 
is $M=4 \times 512=2048$
and the number of energy points is $N_{\rm ene}=1000$. 
The imaginary part of the energy ($z=E+i \gamma$) is set to
$\gamma=0.002$au (=0.0544eV).  
Figure \ref{fig:sCOCG_totalCF} shows the decay behavior of 
the a-RN.
Here and hereafter,
the starting basis $|j \rangle$ for silicon systems
is set to an sp$^3$-hybridized basis on an atom.
We plot four cases with 
different reference energy points.
In result, 
all the cases follow an universal curve till the machine accuracy
and the choice of the reference energy point
does not affect the calculated a-RN.
The behavior within a small iteration ($n \le 70$) 
is also plotted in  Fig.~\ref{fig:total_vs_local}. 
We should note that 
the observed decay at the early stage ($n \le 30$)
is important from practical viewpoint, 
since the iteration number of $n \simeq 30$ 
is enough for the application 
in Sec.~\ref{sec:Si001}.

The convergence behavior was studied also 
in a bulk fcc Cu of 1568 atoms, a metallic system.
The simulation cell is 
a $7\times 7 \times 8$ supercell of the cubic unit cell.
We constructed the Hamiltonian matrix from 
the second-order form ($H^{(2)}$) of the 
tight-binding linear muffin-tin orbital theory.~\cite{OKAndersen.84} 
In Fig.~\ref{fig:total_vs_local}, 
the a-RN is plotted 
with the starting bases of 
atomic s and t$_{2g}$ orbitals. 
Since the present method is a general linear-algebraic theory 
with a short-range Hamiltonian (matrix),
the convergence behavior 
shows no generic difference,
between semiconductor and metal or 
between different starting orbitals.
In fact, 
the curves in Fig.~\ref{fig:total_vs_local}
behave similarly in magnitude  
with a large iteration number,
for example $n \ge 60$, 
though differently 
with a smaller iteration number.

%
%

\subsection{Computational cost within one iteration}
\label{Number_of_operations}

\begin{table}[b]
\begin{tabular}{|p{2.5cm}|p{1.2cm}|p{2.4cm}|p{1.8cm}|} \hline
 Method          & Inner Product     & Scalar-Vector Product & Matrix-Vector Product \\ \hline
 COCG            & $3MN_{\rm ene}$   & $3MN_{\rm ene}$       & $MM_{\rm int} N_{\rm ene}$ \\ \hline
 sCOCG (total)   & $3M$  & $3MN_{\rm ene}$            & $MM_{\rm int}$ \\ \hline
 sCOCG (present) & $3M$  & $3M_{\rm int} (N_{\rm ene} \! - \! 1) +3M$ & $MM_{\rm int}$ \\ \hline
\end{tabular}
\caption{\label{tab:operation}
Numbers of operations within one iteration;
(i) Inner product
in eqs.~(\ref{eq:def_alpha}) and (\ref{eq:def_beta}), 
(ii) Scalar-vector product in 
eqs.~(\ref{eq:xn_def}), (\ref{eq:rn_def}), (\ref{eq:pn_def}) and 
eqs.~(\ref{eq:xn_def_shift}), (\ref{eq:pn_def_shift}), (\ref{eq:shift-res})  
and 
(iii) Matrix-vector product 
in $A{\bm p}_n$ of eq.~(\ref{eq:def_alpha}).
The parameters in the table are as follows;
$M$: 
the dimension of the original Hamiltonian matrix, 
$M_{\rm int}$: 
the number of orbitals within interaction range for one orbital,
$N_{\rm ene}$: 
the number of energy points. 
Here, the cases of the three methods are plotted;
(1)the conventional COCG method, labeled \lq COCG',
(2)the shifted COCG method with the calculation of {\it all} the elements
of the Green's function and the RN, labeled \lq sCOCG (total)'
and (3)the actual procedure 
in the present calculation, labeled \lq sCOCG (present)'.
See the text for details.
}
\end{table}
%

Numbers of operations within one iteration are estimated  
in Table~\ref{tab:operation}.
Two points are found  
for the drastic reduction of  computational  cost,
when the present method is compared 
to the conventional COCG method.
For comparison, 
the case of the conventional COCG method 
is  shown in the column labeled \lq COCG', 
in which 
the conventional COCG methods is applied, independently, 
to all the systems ($N_{\rm ene}$ systems)
and the matrix-vector product governs  the computational cost. 
In the shifted COCG method, on the other hand, 
the computational cost of the matrix-vector product 
is reduced by $1/N_{\rm ene}$ 
($MM_{\rm int}N_{\rm ene} \Rightarrow MM_{\rm int}$),
since the actual matrix-vector product is carried out 
{\it only} for the reference system,
as discussed in Sec.~\ref{sec:shifted_COCG}.
Then the scalar-vector products   
may give a significant contribution to the computation,  
if {\it all} the elements ($M$ elements) of the vectors 
$\bm{x}_{n}^{\sigma}$, $\bm{p}_{n}^{\sigma}$, and $\bm{r}_{n}^{\sigma}$
are calculated for all the systems, 
which 
is shown in the column labeled \lq sCOCG (total)'.
In the present calculation, however,
we need the elements of these vectors 
only within the interaction range ($M_{\rm int}$ elements),
as discussed in Sec.~\ref{SECTION-RN}.
Since the vectors
$\bm{x}_{n}^{\sigma}$, $\bm{p}_{n}^{\sigma}$, and $\bm{r}_{n}^{\sigma}$
in the shifted systems ($N_{\rm ene}-1$ systems)
are updated 
using eqs.~(\ref{eq:xn_def_shift}), (\ref{eq:pn_def_shift}), (\ref{eq:shift-res}),
the  update procedure can be carried out 
only for the necessary elements ($M_{\rm int}$ elements).
The calculation only for these elements gives 
another drastic reduction of the computation cost
($3M(N_{\rm ene}-1) \Rightarrow 3M_{\rm int}(N_{\rm ene}-1)$),
which is shown in the column labeled \lq sCOCG (present)'.

\begin{figure}[t]
\begin{center}
\resizebox{0.52\textwidth}{!}{
  \includegraphics{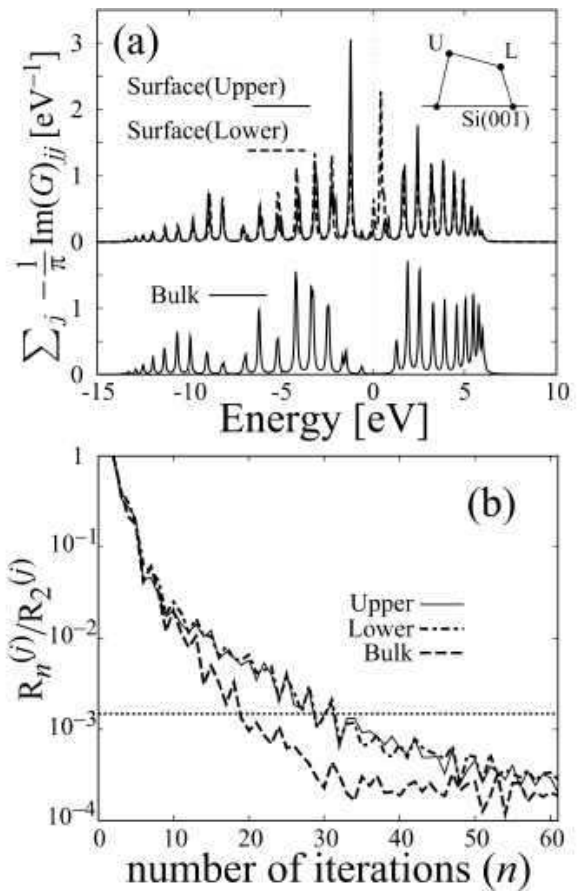}
}
\end{center}
\vspace{-3mm}
\caption{\label{fig:smple1024} 
Result of a slab for Si(001) surface
with asymmetric surface dimers; 
(a) Local density of states   
($\sum_j -(1/\pi){\rm Im}G_{jj}$) 
for surface and bulk atoms,
calculated with the KS dimension of $n=30$. 
The surface atoms are classified into 
the upper and lower atoms of the asymmetric dimer,
illustrated in the inset. 
The \lq bulk' atom is 
an atom on deeper layers of the present slab system. 
(b) Decay behavior of a-RN 
for the atoms that appear in (a).   
The horizontal dotted line is an eyeguide for 
a typical convergence criteria. 
}
\end{figure}

\section{Applications to electronic structure calculation}
\label{sec:example}

\subsection{Reconstruction on Si(001) surface}
\label{sec:Si001}

The MD simulation with the shifted COCG method 
was tested in Si(001) surface reconstruction.
The calculated system is a slab of 1024 atoms 
constituted of 16 layers with 64 atoms on each layer. 
The temperature parameter of the Fermi distribution function in 
eq.~(\ref{eq:rho_int}) is set to $\tau$=0.005au(=0.136eV).
We use the Hamiltonian and energy function in 
Ref.~\cite{kwon-etal.94}. 
Other methodological details are the same as in Sec.~\ref{SECTION-RN}. 
The atomic structure is relaxed, 
with the Hellmann-Feynman force on atoms,
from an appropriate surface atomic configuration
into the ground-state structure.
Since the force on atoms is given
after the calculation of the density matrix,
the simulation was carried out by 
a double-loop iterative procedure;
The inner loop is 
the iterative procedure of the shifted COCG method 
for calculating the density matrix 
with a given atomic configuration or a given Hamiltonian.
The outer loop is the update of the atomic configuration,
with force on atoms, to as to minimize the energy.
When the KS dimension, 
or  the iteration number of the inner loop,
is $n=30$ or larger,  
the resultant surface atoms form asymmetric dimers 
illustrated in the inset of Fig.~\ref{fig:smple1024}(a),
as should do.~\cite{Chadi,Ramstad} 
The resultant tilt angle of the asymmetric dimers 
is $\theta = 13.4^\circ$, 
which agrees with experimental values of $\theta$,
between $5^\circ$  and $19^\circ$.~\cite{Pollmann-etal.96}

Figure~\ref{fig:smple1024}(a) shows the imaginary part 
of the Green's function $\sum_j -(1/\pi){\rm Im}G_{jj}$ 
summed up over the orbitals within a specific atom,
which corresponds to the local density of states (lDOS).
As a general property of the KS method, 
the number of peaks 
in $-(1/\pi){\rm Im}G_{jj}$ equals 
the iteration number or the KS dimension.
When the lDOS of the surface and bulk atoms are compared,
the lDOS of the surface atoms have characteristic peaks 
within $ -1{\rm eV} \le E \le +0.5{\rm eV}$,
because the upper surface atom has 
an {\it occupied} surface state
and the lower one has an {\it unoccupied} surface state. 
The reproduction of these surface states 
is the reason why 
the correct surface reconstruction is reproduced,  
even with a small number of the KS dimension ($n = 30$).

Figure~\ref{fig:smple1024}(b) shows 
the a-RN for these atoms 
as the function of the iteration number
or the KS dimension.
Here the a-RN for an atom is defined
as the average of the a-RN, eq.~(\ref{eq:conv_fact00}),
among the orbitals within the atom.
In result, 
the a-RN for the surface atoms 
decays similarly, 
while that for the bulk atom faster. 
Considering the fact that 
the required number of iterations is $n$=30  
to obtain appropriate surface reconstruction, 
a practical convergence criteria is estimated as 
the horizontal dotted line in Fig.~\ref{fig:smple1024}(b)
on the order of $R_n^{(j)}/R_2^{(j)}\sim 10^{-3}$.
If this convergence criteria is used,
the optimal number of iterations is approximately 18 for the bulk atom,
less than that for the surface atoms ($n = 30$). 
In other words, 
the optimal iteration number is determined 
for {\it microscopic} freedoms or independently among atoms or bases ($|j \rangle$).
Moreover, 
the microscopic control can be carried out 
{\it dynamically}, or at every step in MD simulations. 
In short, 
the observation of the a-RN gives
a definite way of 
controlling the accuracy {\it microscopically and dynamically},
which is important among practical investigations.

\subsection{Metal system: fcc Cu}

\begin{figure}[t]
\begin{center}
 \includegraphics[height=7cm,clip]{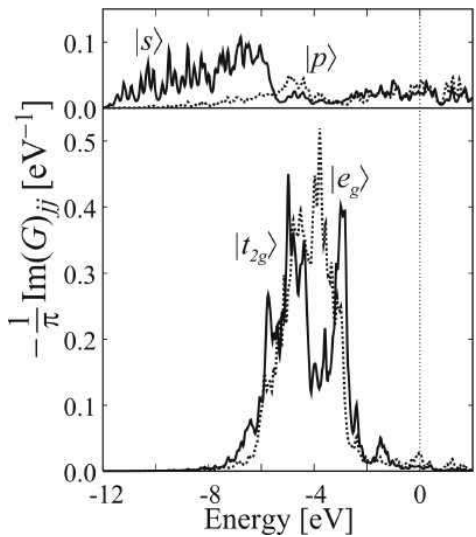}
\end{center}
\vspace{-3mm}
\caption{\label{fig:Cu1568B-1} 
Result of fcc Cu with 1568 atoms;
Partial density of states $-\frac{1}{\pi}{\rm Im}G_{jj}$  
for each orbital. 
}
\end{figure}

We  calculated also the electronic structure of  
a bulk fcc Cu of  1568 atoms.
The  technical details are already explained in Sec.~\ref{SECTION-RN}.
The well-converged partial  DOS is shown in Fig.~\ref{fig:Cu1568B-1} 
for s, p, e$_g$ and t$_{2g}$ orbitals.
The result reproduces 
the essential characteristics, e.g. the resonance behavior of s and p 
orbitals and  the energy separation between e$_g$ and t$_{2g}$ orbitals.

Another important property for analyzing cohesion is 
the crystal orbital Hamiltonian populations (COHP) 
defined as follows;~\cite{COHP}
\begin{eqnarray}
  C_{IJ:\alpha}(E) &=& -\frac{1}{\pi}\sum_{\beta} 
  {\rm Im} \, G_{I\alpha, J\beta}(E + i \gamma) \, H_{J\beta , I\alpha} , \\
  C_{IJ}(E) &=& \sum_{\alpha} C_{IJ:\alpha}(E)  , 
\label{COHP}
\end{eqnarray}
where $I$ and $J$ denote the atomic positions and $\alpha$ and $\beta$ 
orbitals. 
The quantity $C_{IJ}$ is the COHP
and we call the quantity $C_{IJ:\alpha}$ partial COHP (PCOHP).
The energy integration of the COHP (ICOHP)
has the dimension of energy.
The off-site term of ICOHP gives
a quantitative discussion of 
the cohesive mechanism, 
because its negative and positive parts are 
the energy gain and loss in 
the electronic structure energy for cohesion,
respectively.

\begin{figure}[htbp] 
\begin{center}
 \includegraphics[height=11cm,clip]{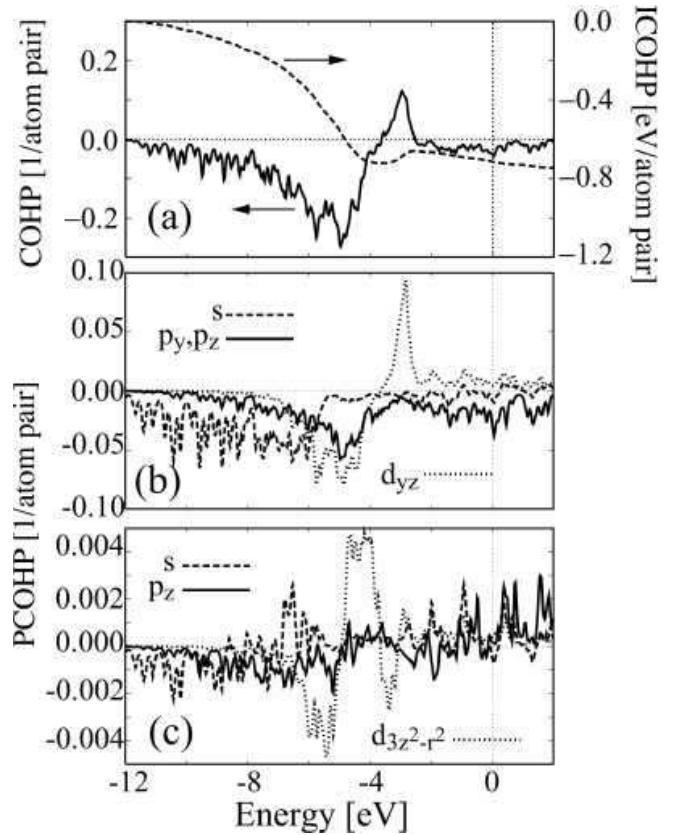}
\end{center}
\vspace{-3mm}
\caption{\label{fig:Cu1568B-2} 
Result of fcc Cu with 1568 atoms;
COHP and ICOHP for a first nearest-neighbor atom pair (a)
and 
PCOHP for a first (b) or second (c) nearest-neighbor atom pair. 
}
\end{figure}

Figure \ref{fig:Cu1568B-2}(a) shows 
the COHP and ICOHP  
for a nearest-neighbor atom pair.  
The PCOHP $C_{IJ:\alpha}(E)$ is also shown in 
Fig.~\ref{fig:Cu1568B-2}(b)
only for the orbitals ($\alpha$) with major contribution. 
Since a nearest-neighbor pair 
lies along $(011)$ direction in fcc, 
the significant values in the PCOHP 
come from  
$\alpha$=s, $({\rm p}_y \pm {\rm p}_z)/\sqrt{2} $ and ${\rm d}_{yz}$ orbitals.
Figures \ref{fig:Cu1568B-2}(a)(b) show that
the two characteristic peaks of the COHP
at $E\simeq -5~{\rm eV}$ and $-3$ eV
are contributed 
mainly by the PCOHP of the ${\rm d}_{yz}$ orbital. 
The negative and positive peaks  originate from 
the bonding and anti-bonding coupling, respectively,  
of the t$_{2g}$ orbitals among the nearest-neighbor atom sites. 
Though the two corresponding peaks can be seen also 
in the PDOS of the t$_{2g}$ orbital in Fig.~\ref{fig:Cu1568B-1},
the COHP, unlike PDOS,  informs us 
the bonding or anti-bonding character 
of the corresponding state. 
From Fig.~\ref{fig:Cu1568B-2}(b),
we found that contributions from s and p orbitals are also appreciable. 
Moreover, the PCOHP for a {\it second} nearest-neighbor pair 
is plotted in Fig.~\ref{fig:Cu1568B-2}(c).
Since a second nearest-neighbor pair 
lies along $(001)$ direction, 
the significant values in the PCOHP come from  
s, ${\rm p}_z$ and ${\rm d}_{3z^2-r^2}$ orbitals,
though their magnitude is one order  smaller than those for the first neighbor pair. 

The present analysis demonstrates that, 
since the shifted COCG method 
can give the Green's  function with the machine accuracy,
the resultant spectra reproduce
the correct cohesive mechanism,
in which the role of each orbital is well described 
not only for the major contribution from the first nearest-neighbor coupling
but also for a minor contribution from the second nearest-neighbor coupling.

\section{Conclusive discussion}
\label{sec:conclusion}

In the present paper,  
we introduced the shifted COCG method based on the Krylov subspace
and used the method as an iterative solver algorithm of the Green's function
in large electron systems.
We analyzed  
the convergence behavior by means of the a-RN,
which establishes 
a definite way of  controlling accuracy. 
The theory realizes a practical method not only for MD simulations 
but also for obtaining the fine-resolution spectra, 
such as (P)DOS and COHP, without calculating eigen states. 
The method was applied 
to semiconductor and metal 
and the above statements were confirmed numerically.

When the present method is compared with the SD-KS method, 
we conclude that
the two KS methods are complementary in the practical viewpoint 
and we should choose one, according to the purpose;
If one would like to obtain 
the Green's function in a {\it very fine} energy resolution,
the shifted COCG method should be used,
because it can reach the machine accuracy. 
On the other hand, 
as is discussed in Appendix~\ref{sec:CEbySD},
the SD-KS method is suitable 
for obtaining the density matrix 
without numerical energy integration of the Green's function, 
which is a typical situation in MD simulations.

Finally, we point out the generality of the present theory.
Since the shifted COCG method is based 
on a general linear-algebraic theory with large matrices,
it is applicable not only 
to electronic structure calculation with atomic orbital bases 
but also to calculation with other bases.
Moreover, the method may be useful
in many theoretical fields other than electronic structure theory,
if a theory is reduced to a set of shifted linear equations.

\section*{Acknowledgments}
Computation was carried out 
at the Center for Promotion of Computational Science and Engineering (CCSE) of 
Japan Atomic Energy Research Institute (JAERI) 
and also partially carried out at 
the Supercomputer Center, Institute for Solid State Physics, University of Tokyo. 
This work is financially supported by Grant-in-Aid 
from the Ministry of Education, Culture, Sports, Science and Technology 
and also by ACT-JST and CREST-JST.

\appendix
\begin{figure}[bhtp] 
\begin{center}
\resizebox{0.48\textwidth}{!}{
  \includegraphics[width=17cm,clip]{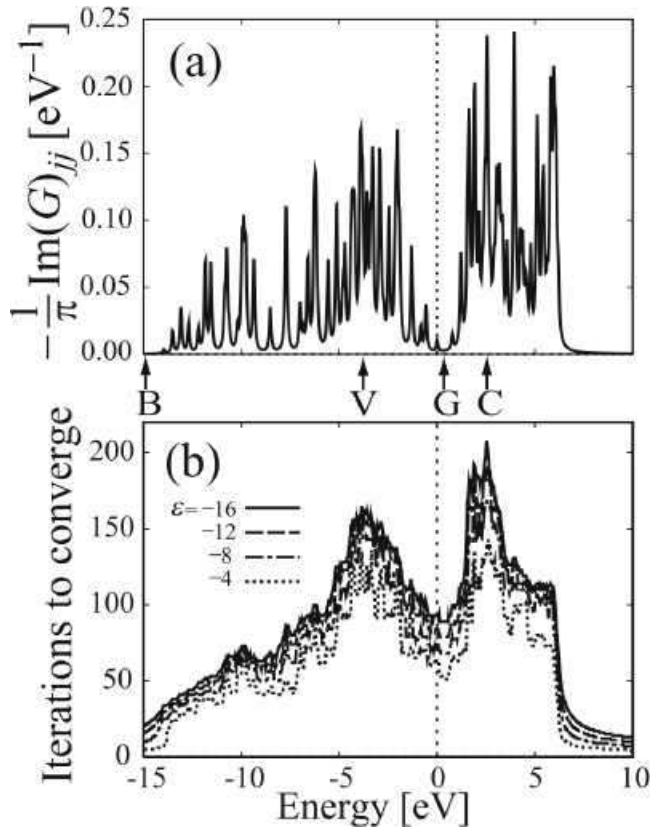}
}
\end{center}
\vspace{-2mm}
\caption{\label{fig:cocgs01ldos0} 
(a) Well-converged Green's function 
and (b) convergence behavior at different energy points.
The calculation was carried out  
with Si 512 atoms and
the top of the valence band locates at $E=0~{\rm eV}$;
 (a) Imaginary part of the well-converged Green's function 
 ($-\frac{1}{\pi}{\rm Im}G_{jj}(E+i\gamma)$) with $\gamma=0.002$au. 
 (b) The iteration numbers  to fulfill 
 the converge criterion 
$||\bm{r}_{n}^{(j)}||^2 < 10^{\varepsilon}$
with $\varepsilon = -4$ to $-16$. 
Note that four energy points are picked out,
at the band bottom (B), 
in the valence band (V), 
in the gap (G) and in the conduction band (C),
for discussion (see text). 
}
\end{figure}%
%

 \begin{figure}[htbp] 
 \begin{center}
 \resizebox{0.48\textwidth}{!}{
 \includegraphics[width=6cm,clip]{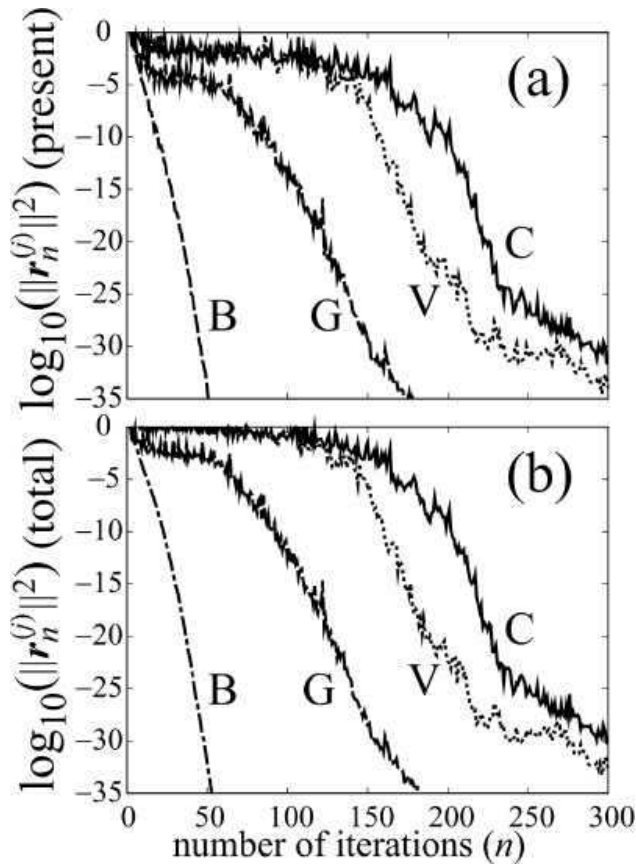}
 }
 \end{center}
 \caption{\label{fig:eps_cocgs01-RN} 
 Iteration dependence of (a) the  residual norm (RN) 
 in the present method 
 and 
(b) the \lq total' residual norm (t-RN).
The symbols \lq B', \lq V', \lq G'  and \lq C' indicate 
the energy points that are shown by arrows in 
Fig.~\ref{fig:cocgs01ldos0}~(a). 
}
\end{figure}%
%

\section{Numerical aspects in Krylov-subspace methods}
\label{sec:appendix}

This appendix is devoted to  
several aspects of the two KS methods
(i) the shifted COCG method, the main subject of this paper,
and 
(ii) the SD-KS method~\cite{takayama_etal.04}. 
Particularly, 
numerical aspects will be discussed,
including robustness against round-off error.  
Examples are demonstrated with silicon crystals, 
as in Sec.~\ref{RN_and_Opera}. 

\subsection{Shifted COCG method}
\label{sec:form_shifted_COCG}

For the shifted COCG method,
we discuss 
(I) the convergence behavior among different energy points and
(II) the convergence behavior of long-distance component
of the Green's function. 
These discussions clarify how and why the present method 
is so successful. 

First, 
we used the conventional COCG method at all the energy points
($N_{\rm ene}=1000$ points),
so as to investigate the convergence behavior 
among different energy points. 
Figure \ref{fig:cocgs01ldos0}(a) shows the imaginary 
part of the well-converged Green's function, 
corresponding to the density of states (DOS).   
The spectrum consists of a set of spikes
with a finite width, 
owing to a small imaginary part of energy
($\gamma=0.002~{\rm au}$=0.0544eV). 
The required iteration numbers with various convergence criteria 
are seen in Fig.~\ref{fig:cocgs01ldos0}(b),
in which the convergence criterion is set
to be $||\bm{r}_{n}^{(j)}||^2 < 10^{\varepsilon}$
with $\varepsilon = -4$ to $-16$.
The resultant DOS profiles
among these criteria
are indistinguishable
from that of Fig.~\ref{fig:cocgs01ldos0}(a).
Figures~\ref{fig:cocgs01ldos0}(a) and (b) indicate that  
an energy point with a larger value of DOS requires 
a larger number of iterations, 
because the KS dimension should be larger 
in order to distinguish individuals among densely distributed nearby states.

In Fig.~\ref{fig:eps_cocgs01-RN}~(a),
the decay behavior of the RN is plotted  
for the four chosen energy points 
that were already discussed in 
Fig.~\ref{fig:sCOCG_totalCF} and Fig.~\ref{fig:cocgs01ldos0}~(a).
When Fig.~\ref{fig:eps_cocgs01-RN}~(a) and 
Fig.~\ref{fig:sCOCG_totalCF} are compared,
one can see that 
the decay behavior of the RN 
are quite different among the four energy points
but the a-RN is universal, 
as should be from eq.~(\ref{eq:shift-res}). 
For example, 
we pick out the case 
in which the reference energy is chosen 
as the point labeled by (B).
In the case, 
the RN $|| {\bm r}_n^{(j)} ||^2$ and 
the shift coefficient $\pi_{n}^{\sigma}$ 
go to the extreme orders of $10^{- 250}$ and $10^{+ 250}$, 
respectively, at $n=300$.
Even in such an extreme case, 
the computational procedure works well and 
the a-RN follows the universal curve, 
as in Fig.~\ref{fig:sCOCG_totalCF}.

Second, we discuss 
the convergence behavior of the Green's function 
including its {\it long-distance} component,
unlike in Sec.~\ref{RN_and_Opera}.
For monitoring the convergence behavior,  
we should define the \lq total' residual norm (t-RN), 
instead of the RN in eq.~(\ref{eq:ref-p-RN}), as
\begin{eqnarray}
|| {\bm r}_n^{(j)} ||^2 &=& \sum_i^{M} |\langle i| r_n^{(j)} \rangle |^2,
   \label{eq:ref-t-RN} 
\end{eqnarray}
where the elements are summed up 
among {\it all} the bases ($M$ bases).
In other words, the t-RN shows 
the convergence behavior for 
{\it all} the elements of the Green's function ($G_{ij}$ or $G(\bm{r},\bm{r}')$),
including its long-distance components. 
We should recall that
the RN in eq.~(\ref{eq:ref-p-RN})  
is defined only for the short-distance components. 
The results are shown in Fig.~\ref{fig:eps_cocgs01-RN} (b) ,
in which the decay behavior 
is slower than that in Fig.~\ref{fig:eps_cocgs01-RN} (a) 
at an earlier stage ($n \le 30$), 
though the behavior is the same at later stages.  
The difference at the earlier stage appears, because 
the accurate description of the Green's function
at further distances needs a larger number of KS bases 
or a larger iteration number.  
Moreover, 
the computational cost is enormous 
to calculate {\it all} the elements of the Green's function, 
as shown in the \lq sCOCG(total)' case 
of Table~\ref{tab:operation}.
The present discussion clarifies
that only the short-distance components of the Green's function
are required and calculated in practical applications.

\subsection{Subspace diagonalization method}
\label{sec:CEbySD}

Here 
the subspace diagonalization method 
based on the KS (SD-KS method)\cite{takayama_etal.04}
is discussed for 
the comparison with the shifted COCG method. 
Though the two methods, commonly,  
give the density matrix or Green's function within the KS, 
the difference between them comes
from the computational cost and 
the effect of numerical round-off error.

The practical procedure of the SD-KS method 
is summarized as follows;
an orthogonal basis set for the KS is constructed
by the Lanczos process, a three-term recurrence relation;
$\{ | K_1^{(j)} \rangle \equiv | j \rangle, 
 | K_2^{(j)} \rangle,... | K_{\nu}^{(j)} \rangle \}$.~\cite{takayama_etal.04}
Here the number of bases $\nu$ is 
the dimension of the KS, $K_\nu(H,| j \rangle)$. 
This process creates simultaneously 
the reduced Hamiltonian matrix $H^{K(j)}$ within the KS
$((H^{K(j)})_{nm} \equiv \langle K_n^{(j)} | H  | K_m^{(j)} \rangle)$,
as a small tridiagonal matrix.  
Then the reduced matrix $H^{K(j)}$ is diagonalized 
and we obtain the eigen values $\varepsilon_{\alpha}^{(j)}$ 
and the coefficients $C_{\alpha n}^{(j)}$ ($\alpha = 1,2,....\nu$),
where the eigen vectors are given as
$| w_{\alpha}^{(j)} \rangle \equiv 
\sum_{n=1}^\nu  C_{\alpha n}^{(j)}  | K_{n}^{(j)} \rangle$. 
The Green's function is given as 
\begin{eqnarray}
\langle i | G | j \rangle \Rightarrow 
\sum_n^{\nu}  \langle i |  K_{n}^{(j)} \rangle 
\langle  K_{n}^{(j)} | {G}_{\nu}^{(j)} | j \rangle
\end{eqnarray}
with the definition of
\begin{eqnarray}
 {G}_{\nu}^{(j)}(z) 
&\equiv& \sum_{\alpha}^{\nu_{}} 
\frac{\mid w_{\alpha}^{(j)} \rangle \langle w_{\alpha}^{(j)} \mid }{z-\varepsilon_{\alpha}^{(j)} } .
\label{eq:A5}
\end{eqnarray}
The density matrix 
can be given in a similar manner. ~\cite{takayama_etal.04} 
The calculated  band structure energy 
shows a rapid convergence as the function of 
the number of the KS bases $\nu$,
both in semiconductor ~\cite{takayama_etal.04}  and 
metal (fcc Cu, unpublished). 
In both cases, the result is well converged, typically,  
with $\nu=30$ . 
We have simulated the reconstruction on Si(001) surface,\cite{takayama_etal.04} 
as in Sec.~\ref{sec:Si001},
and the resultant atomic position or lDOS 
agrees with the ones obtained 
by the shifted COCG method. 

Several differences in numerical treatment are found
between the SD-KS method and the shifted COCG method.
The SD-KS method gives
the Green's function analytically in eq.(\ref{eq:A5})
and its energy integration for the density matrix,
eq~(\ref{eq:rho_int}), 
can be given also analytically, 
while the shifted COCG method gives 
the Green's function numerically on 
a given set of energy points and 
its energy integration should be carried out
with a careful observation of the numerical error. 
Moreover,
the computational cost of the SD-KS method is smaller than,
typically a half of,  
that of the shifted COCG method,
because the SD-KS method requires only real vectors, such as 
$|  K_{n}^{(j)} \rangle, | w_{\alpha}^{(j)} \rangle$,
while the shifted COCG method requires
several complex vectors, such as  $| x_{n}^{(j)} \rangle, | r_{n}^{(j)} \rangle $.

Hereafter we discuss a crucial difference of the SD-KS method 
from the shifted COCG method; 
the SD-KS method
shows a numerical instability with a very large number of KS bases ($\nu$), 
owing to the accumulation of round-off error. 
The above instability is analyzed 
by introducing the RN 
with eq.~(\ref{eq:residual_G0}).
An element of the RN is given as
\begin{eqnarray}
\langle i | r_\nu^{(j)} \rangle  &=& 
  \langle i | I-(z-H)  {G}_\nu^{(j)}(z) | j \rangle   \nonumber  \\
&=& \langle i | j \rangle - \sum_{\alpha}^\nu 
  \frac{\langle i | z-H |  w_{\alpha}^{(j)} \rangle \langle  w_{\alpha}^{(j)} | j \rangle}
  {z-\varepsilon_{\alpha}^{(j)} }  \nonumber \\
&=& \langle i | j \rangle - \sum_{\alpha, n =1 }^\nu \left\{ 
  \frac{\langle i | z-H | K_{n}^{(j)} \rangle C_{\alpha n}^{(j)} }
  {z-\varepsilon_{\alpha}^{(j)} } \right. \nonumber \\
&& \times  \left.  \sum_{m=1}^\nu 
 C_{\alpha m}^{(j)} \langle K_{m}^{(j)} | j \rangle \right\}.
\label{eq:A6}
\end{eqnarray}
This quantity can be calculated with a negligible computational cost, 
since all the quantities in eq.~(\ref{eq:A6}), 
that is $ \{ \varepsilon_{\alpha}^{(j)} \}$, 
$\{ C_{\alpha n}^{(j)} \}$, 
$\{ \langle i | K_{n}^{(j)} \rangle\}$ and 
$\{  \langle i | H|K_{n}^{(j)} \rangle \}$, 
are always calculated in the generating procedure
of the Green's function ${G}_\nu^{(j)}(z)$. 
The a-RN $R_{\nu}^{{\rm } (j)}$ 
can be defined in a similar way as in eq.~(\ref{eq:conv_fact00}). 
The a-RN was examined 
for Si crystal in different system sizes
(512, 4096, and 32768 atoms),
and the results are shown in Fig.~\ref{fig:SDfomlog2R2sc}. 
Here a problematic situation appear in the cases of 512 and 4096 atoms,
because the a-RN begins to grow 
after an appropriate value of the KS dimension $\nu$.
So as to analyze the growth of error, 
the RN with the case of 512 atoms 
is shown in Fig.~{\ref{fig:SDpiR22} with its energy dependence. 
The spectrum consists of only spikes 
before the  growth of error 
($\nu=30$, Fig.~{\ref{fig:SDpiR22}(a)), 
while a finite background appears 
after the beginning of the growth of error 
($\nu=100$, Fig.~{\ref{fig:SDpiR22}(b)).

\begin{figure}[htbp] 
\begin{center}
\resizebox{0.48\textwidth}{!}{
  \includegraphics{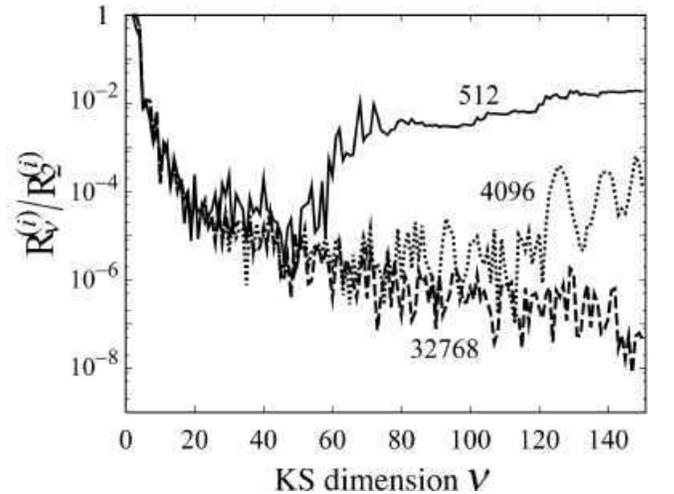}
}
\end{center}
\vspace{-5mm}
\caption{\label{fig:SDfomlog2R2sc} 
Decay behavior of the a-RN 
in the SD-KS method 
for Si crystals with 512, 4096 and  32768 atoms. 
}
\end{figure}

\begin{figure}[h]
\begin{center}
\centerline{
 \includegraphics[width=8.5cm,clip]{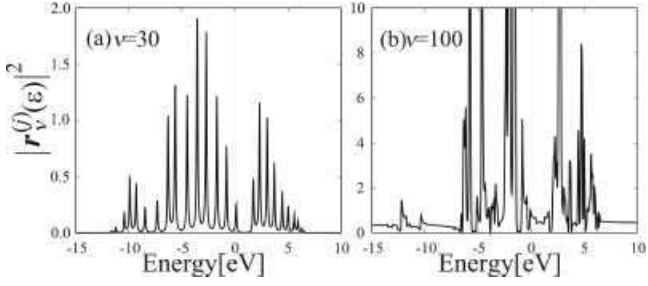}
}
\end{center}
\vspace{-5mm}
\caption{\label{fig:SDpiR22} 
Energy dependence of the RN 
in the subspace diagonalization method 
for Si crystal with 512 atoms. 
The KS dimension is (a) $\nu=30$ and (b) $\nu=100$. 
}
\end{figure}%

The growth of error occurs, 
because the loss of orthogonality 
($  \langle K_n^{(j)} | K_m^{(j)} \rangle \ne \delta_{nm}$) 
always happens in actual calculation after a long iteration
of the Lanczos procedure, 
owing to the accumulation of round-off error. 
In such a case, 
the calculated  eigen vectors $|w_{\alpha}^{(j)} \rangle$  
have a finite deviation $|\xi_{\alpha}^{(j)} \rangle$ from the true eigen vectors;
\begin{equation}
\label{eq:sd_eig_eq}
H^{K(j)}|w_{\alpha}^{(j)} \rangle = \varepsilon_{\alpha} |w_{\alpha}^{(j)} \rangle 
+ |\xi_{\alpha}^{(j)} \rangle. 
\end{equation}
Using eqs.(\ref{eq:A6}) and (\ref{eq:sd_eig_eq}),
we obtain
\begin{eqnarray}
\langle i|r_\nu^{(j)}\rangle &=& 
\langle i | I- (z-H)  {G}_{\nu}^{(j)}(z) | j \rangle 
\nonumber \\
&=& \langle i | j \rangle - \sum_{\alpha}^\nu 
  \frac{\langle i | z-H |  w_{\alpha}^{(j)} \rangle \langle  w_{\alpha}^{(j)} | j \rangle}
  {z-\varepsilon_{\alpha}^{(j)} }  \nonumber \\
&=&  \langle i | d \rangle  \nonumber \\
& &+ \sum_{\alpha =1}^{\nu} 
\left\{ 
\langle i |  \xi_{\alpha}^{(j)} \rangle  
+\langle i | \delta H^{K(j)}  | w_{\alpha}^{(j)} \rangle 
\right\}
\frac{\langle w_{\alpha}^{(j)} | j \rangle}{z-\varepsilon_{\alpha}^{(j)}}, \nonumber \\
& & \label{eq:geta}
\end{eqnarray}
where 
we define $\delta H^{K(j)} \equiv H - H^{K(j)}$ and use
\begin{eqnarray}
|d \rangle 
 &\equiv& |j \rangle -\sum_{\alpha=1}^\nu 
  |w_\alpha^{(j)}\rangle \langle w_\alpha^{(j)}| j \rangle  \nonumber \\
 &=& \sum_{n=2}^\nu |K_n^{(j)}\rangle \langle K_n^{(j)}|  K_1^{(j)} \rangle .
 \label{eq:dev-vec}
\end{eqnarray}
The last equality of eq.~(\ref{eq:dev-vec}) 
is given by 
\begin{eqnarray}
\sum_{\alpha=1}^\nu 
  |w_\alpha^{(j)}\rangle \langle w_\alpha^{(j)}|  = 
  \sum_{n=1}^\nu |K_n^{(j)}\rangle \langle K_n^{(j)}| 
\end{eqnarray}
and  $ |K_1^{(j)}\rangle \equiv |j \rangle$. 
The first and second terms in eq.~(\ref{eq:geta}) 
correspond to the finite background and the spiles
in Fig.~\ref{fig:SDpiR22}(b), respectively. 
With a small number of bases, as in Fig.~\ref{fig:SDpiR22}(a), 
the orthogonality between the bases holds exactly
and we obtain $ |d\rangle = |  \xi_{\alpha}^{(j)} \rangle =0$.
Therefore, the energy-independent term, the first term of eq.~(\ref{eq:geta})
does not appear. 
The spikes in Fig.~\ref{fig:SDpiR22}(a) appear,
 because the reduced Hamiltonian matrix in the KS is deviated  
 from the original one 
($\delta H^{K(j)} \equiv H - H^{K(j)} \ne0$).



\begin{thebibliography}{10}
\bibitem{Note_Frommer}
The original paper~\cite{Frommer.02} gives the formulation 
for a general non-Hermitian matrix $A$,
which is called shifted Bi-CGStab algorithm.
In the present context, however,
the matrix $A$ is  complex symmetric and 
the present formulation of the shifted COCG algorithm can be adopted.
\end{thebibliography}
\end{document}